\documentclass[aps,prl,twocolumn,showpacs]{revtex4}
\usepackage{epsfig}
\newcommand{\be}{\begin{equation}} 
\newcommand{\ee}{\end{equation}}
\newcommand{\bea}{\begin{eqnarray}} 
\newcommand{\eea}{\end{eqnarray}}
 
\newcommand{\vp}{{\vec p}}

\begin{document}
\title{Dissipation and elliptic flow at RHIC}
\author{D\'enes Moln\'ar}
\affiliation{Department of Physics, Ohio State University, 174 West 18th Ave,
Columbus, Ohio 43210, USA}
\author{Pasi Huovinen}
\affiliation{School of Physics and Astronomy, University of Minnesota,
              Minneapolis, Minnesota 55455, USA}
\affiliation{Helsinki Institute of Physics, P.O.\ Box 64,
              FIN-00014 University of Helsinki, Finland}
 \affiliation{Department of Physics, P.O.\ Box 35,
              FIN-40014 University of Jyv\"askyl\"a, Finland}
\date{\today}
\begin{abstract}
We compare elliptic flow evolution from ideal hydrodynamics and 
covariant parton transport theory,
and show that, for conditions expected at RHIC, 
dissipation significantly reduces elliptic flow 
even for extreme parton cross sections
and/or densities 
$\sigma_{gg} \times dN/d\eta(b{=}0) \sim 45\ {\rm mb } \times 1000$.
The difference between transport and hydrodynamic elliptic
flow is established rather early during the evolution of the system,
but the buildup of elliptic flow is surprisingly insensitive to the
choice of the initial (formation or thermalization) time in both models.
\end{abstract}

\maketitle

{\em Introduction.}
During the early stage of ultra-relativistic heavy ion collisions at
the Relativistic Heavy Ion Collider (RHIC) and the future Large Hadron
Collider (LHC), a deconfined phase called the quark-gluon plasma (QGP)
is expected to be formed.  One way to infer the properties of this
dense partonic phase is to study its collective behavior.

An important experimental probe of collective dynamics 
in noncentral $A+A$ reactions
is elliptic flow, $v_2 \equiv\langle \cos(2\phi)\rangle$, 
the second Fourier moment of the azimuthal momentum distribution 
\cite{flow-review}.
Recent data from RHIC for Au+Au at $\sqrt{s_{NN}} \sim 130-200$ GeV
show a large anisotropy of particle production in the transverse plane
with $v_2$ values reaching up to $0.2$
and saturating in the transverse momentum region $2 < p_\perp < 6$ GeV 
\cite{PHENIXv2,STARv2}.

For $p_\perp < 2$ GeV, i.e., for the bulk of the particles produced at RHIC,
the elliptic flow data can be reproduced from ideal (Euler) hydrodynamics. 
In this completely
nondissipative theory elliptic flow is sensitive to the equation of 
state (EOS) of dense nuclear matter 
\cite{Ollitrault,Huovinen:2001cy, Kolb:2000sd,Teaneyhydro}.
It is remarkable that agreement with the data requires
\cite{Huovinen:2001cy, Kolb:2000sd,Teaneyhydro} an EOS 
with deconfinement phase transition,
providing one of the strongest arguments for QGP formation at RHIC.

Despite its successes, 
the assumption of no dissipation has limitations.
For example, ideal hydrodynamics
fails~\cite{snellings} to saturate elliptic flow for $p_\perp > 2$ GeV,
and overpredicts~\cite{hydroHBT} the ``long'' and ``out''
pion Hanbury-Brown and Twiss (HBT) interferometry radii.
Both of these shortcomings are likely due to the 
neglect of dissipation~\cite{Molnar:v2,Adrianviscos,Teaneyviscos,MolnarHBT},
caused by nonzero mean free paths.
In any case, the mere \emph{proof} that the effect of dissipation is
negligible (for some observable) already 
requires a framework that \emph{allows} for dissipation.

For systems near equilibrium, dissipative corrections can be studied
via Navier-Stokes (viscous) hydrodynamics. In this theory, the
evolution is determined by the equation of state, the shear and bulk
viscosities, and the heat conductivity.  Even though a general
Lorentz-covariant formulation exists, only 1+1D relativistic
solutions are known~\cite{Muronga,Bin:Et}.  Nevertheless, recent
estimates~\cite{Adrianviscos,Teaneyviscos,Muronga} find considerable
viscous corrections to the evolution of the system, and to spectra, elliptic
flow $v_2(p_\perp)$, and pion HBT parameters at RHIC.

To study dissipative effects arbitrarily far from equilibrium,
one can utilize covariant parton transport theory
\cite{ZPC,Bin:Et,ZPCv2,nonequil,inelv2,Molnar:v2,MolnarHBT}.
In this approach elliptic flow depends mainly on
the effective scattering cross section of partons produced
in the collision \cite{ZPCv2,Molnar:v2}. 
Both the magnitude and the saturation feature of elliptic flow at RHIC
can be reproduced~\cite{Molnar:v2} if very large ($\sim 45$~mb)
elastic parton-parton cross sections are assumed.
It is puzzling that these values are an order of magnitude above 
conventional perturbative QCD (pQCD) estimates.
There are, however, recent proposals to alleviate
this problem~\cite{Molnarcoal,Shuryak:2003ty}.

A popular interpretation of the joint success of ideal hydrodynamics
and parton kinetic theory at RHIC is that with $\sim 45$ mb elastic
parton-parton cross sections the opacities are large enough to reach
the ideal hydrodynamic limit, at least at $p_T<2$ GeV.
However, that conclusion ignores that
calculations based on the two theories corresponded to different
initial conditions and thermodynamic properties.

In this paper we provide a systematic comparison of ideal
hydrodynamics and transport theory and investigate the importance of
dissipative effects at RHIC.  Because a thermodynamically consistent
microscopic description of the QGP phase transition is still a
difficult open problem, we gain insight via studying an ideal gas of
massless gluons.

Precursors of this study in Refs.~\cite{Bin:Et,nonequil} analyzed the
transverse energy of particles and found $\sim 20$\% dissipative
corrections.  Here we focus on differential elliptic flow and show
that it is more sensitive to dissipation.

{\em Ideal hydrodynamics and covariant transport theory.}
Ideal (Euler) hydrodynamics is a 
convenient Lorentz-covariant dynamical
theory formulated in terms of macroscopic quantities.
In the context of heavy ion collisions,
its main advantage is the ability to treat phase transitions,
while its main limitation is that it is only applicable 
to systems in local kinetic equilibrium.

The hydrodynamical equations of motion are the local conservation laws
of energy-momentum and net charge
\be \partial_\mu T^{\mu\nu}(x) = 0, \qquad \partial_\mu N_c^\mu(x) = 0
\label{hydroeq}
\ee 
[$x\equiv (t, \vec x)$ is the Minkowski four-coordinate]. 
In ideal hydrodynamics the energy-momentum tensor is assumed to be
that of an ideal fluid,
$T^{\mu\nu} = (\epsilon + p)u^\mu u^\nu - p g^{\mu\nu}$,
where $\epsilon(x)$, $p(x)$, and $u^\mu(x)$ 
are the local energy density, pressure, and flow velocity.
The charge current is related to the local charge density $n_c$ via
$N_c^\mu = n_c u^\mu$.
Once the equation of state $p(\epsilon,n_c)$ is 
specified, Eqs.~(\ref{hydroeq}) can be utilized to follow the 
evolution of the system from any given initial state in local kinetic
equilibrium.

For heavy-ion collision applications, hydrodynamics has to be supplemented
with a freezeout description because the assumption of local equilibrium
breaks down as the local mean free path $\lambda = 1/[\sigma(s) n(x)]$
becomes comparable to the Hubble radius during the expansion. A common
approach is to assume that microscopic scattering rates
$\Gamma_{sc}(T,n) \sim \sigma(T) n$ drop so quickly that one can
consider a \emph{sudden} transition from local equilibrium to a
noninteracting gas. In this case, the fluid is converted to particles
on a 3D spacetime hypersurface via the Cooper-Frye
formula~\cite{CooperFrye}. For systems in chemical equilibrium, a
constant temperature (or constant energy density) hypersurface is
usually chosen because the only relevant parameter is the temperature,
$\Gamma_{sc}(T, n(T))$.

Covariant parton transport theory, in contrast to hydrodynamics, is
based on microscopic quantities, namely, the phasespace distributions
of (quasi)particles and microscopic transition probabilities.  The
main advantage of this approach is that it is applicable out of
equilibrium and models freezeout self-consistently.  However, it
cannot describe phase transitions (without coupling to classical
fields).

We consider here, as in Refs. 
\cite{ZPC,Bin:Et,ZPCv2,nonequil,Molnar:v2,MolnarHBT},
the simplest but nonlinear
form of Lorentz-covariant Boltzmann transport theory
in which the on-shell phase space density $f(x,\vp)$,
evolves with an elastic $2\to 2$ rate as
\bea
p_1^\mu \partial_\mu f_1 &=& S(x, \vp_1) 
+ \frac{1}{16\pi^2} \int\limits_2\!\!\!\!
\int\limits_3\!\!\!\!
\int\limits_4\!\!
\left(f_3 f_4 - f_1 f_2\right)
\left|{\overline{\cal M}}_{12\to 34}\right|^2 \nonumber\\
&&\qquad\qquad\qquad\qquad\quad\times \  \delta^4(p_1{+}p_2{-}p_3{-}p_4)
 \ .
\label{Boltzmann_eq}
\eea
Here $|\overline {\cal M}|^2$ is the polarization averaged scattering matrix 
element squared,
the integrals are shorthands
for $\int_i \equiv \int d^3 p_i / (2E_i)$,
while $f_j \equiv f(x, \vp_j)$.
The initial conditions are specified by the source function $S$.
For our applications below,
we interpret  $f(x,\vp)$ as describing
an ultrarelativistic massless gluon gas with $g=16$ degrees of freedom
(8 colors, 2 helicities).

Eq.~(\ref{Boltzmann_eq}), which corresponds to Boltzmann statistics,
could in principle be extended for bosons
and/or for inelastic processes, such as $gg\leftrightarrow ggg$.
However, no \emph{practical} covariant algorithm yet exists
that can handle, at the opacities expected at RHIC, 
the new nonlinearities these extensions introduce.
We therefore limit our study to quadratic dependence of the collision
integral on $f$.

Boltzmann's \emph{$H$-theorem} states that
Eq.~(\ref{Boltzmann_eq}) drives the system towards a fixed point,
global equilibrium.
In the \emph{hydrodynamic limit}, i.e., when $|{\cal M}|^2\to \infty$,
the transport evolution approaches the ideal hydrodynamic evolution, 
$f(x,\vp) = g \exp[(\mu(x) - p_\nu u^\nu(x))/T(x)]/(2\pi)^3$,
provided hydrodynamics is \emph{not} frozen out via some prescription.
However, for a finite matrix element, an {\em expanding} system, 
if it ever equilibrates, sooner or later evolves out of equilibrium
(except for very special examples~\cite{CsorgoFree}).

{\em Numerical results.}
We study elliptic flow in a typical midperipheral Au+Au collision at
RHIC with impact parameter $b=8$~fm.  The initial conditions were
taken from \cite{Molnar:v2}: a longitudinally boost invariant Bjorken
tube in local kinetic equilibrium at temperature $T_0$ at proper time
$\tau_0$~fm/$c$, with a transverse gluon density distribution that is
proportional to the binary collision distribution for two Woods-Saxon
nuclei.  A thermal momentum distribution is necessary because
hydrodynamics is limited to such initial conditions.  Unlike in usual
hydrodynamic calculations, the initial temperature was the same in the
entire system even though the density of particles was not.  This means
that the system was not in chemical equilibrium.  Motivated by
parton-hadron duality\cite{EKRT}, the initial gluon density was normalized to
$dN_g/d\eta(b=0)=1000$ to match the observed $dN_{ch}/d\eta \sim 600$.
Accordingly, $dN_g/d\eta \approx 250$ for $b=8$~fm.

We solved Eq. (\ref{Boltzmann_eq}) via the MPC algorithm\cite{MPC},
which utilizes parton subdivision\cite{ZPC,nonequil} 
to maintain Lorentz covariance.
The elastic parton cross section was isotropic with
$\sigma_{tot}=3$, $8$, and $20$~mb, which via transport opacity 
scaling~\cite{Molnar:v2} is approximately
equivalent to using a Debye-screened leading-order pQCD $gg\to gg$ cross 
section 
$d \sigma/dt \propto 1/(t-\mu^2)^2$ with $\sigma_{tot} \approx 7$, $19$, 
and $47$ mb.
Parton subdivisions of 180 were used to stabilize the numerical results.

The hydrodynamic solutions were obtained via the code used in
Refs.~\cite{Huovinen:2001cy,Kolb:2001qz}.
Particle number was conserved explicitly
because we are comparing to transport theory with 
elastic $2\to 2$ interactions.
For the massless parton gas considered here, the EOS is $\epsilon = 3p$.
Because the cross section is energy-independent,
freezeout was chosen to occur at a proper density of
$n_g = 0.365$ fm$^{-3}$. This corresponds to the gluon density of a chemically
equilibrated gluon gas at $T=120$ MeV.

\begin{figure}[htpb]
\centerline{\epsfig{file=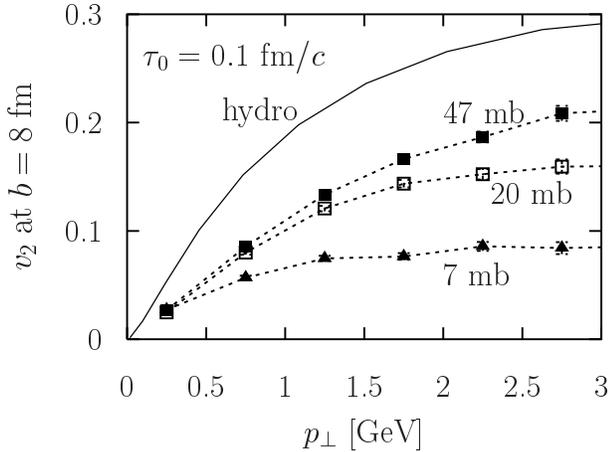,width=3.23in,height=2.38in,angle=0}}
\caption{Comparison of $v_2(p_\perp)$ from ideal hydrodynamics (solid curve) 
and transport theory as a function of the parton cross section (dashed curves),
for $\tau_0=0.1$ fm/$c$.}
\label{fig:1}
\end{figure}
Figure~\ref{fig:1} shows elliptic flow $v_2(p_\perp)$ at freezeout for
$T_0 = 700$~MeV and $\tau_0 = 0.1$~fm$/c$ 
(which are the parameters used in \cite{Molnar:v2}).
From identical initial conditions and thermodynamic properties 
($\epsilon = 3p$ EOS)
ideal hydrodynamics produces a much larger anisotropy than transport theory
for all three parton cross sections $\sigma_{gg} = 7$, $19$, and $47$~mb.
Even for the largest $\sim 45$~mb parton cross section studied,
dissipation reduces $v_2$ by $30-40$\%.
This effect is larger by a factor 1.5--2 than what
earlier studies~\cite{Bin:Et,nonequil} found 
for the final transverse energy $dE_T/dy$
\footnote{See, e.g., Fig.~1 in \protect\cite{nonequil}. The open squares there
correspond to the
$\sigma_{gg}=47$ mb case in this study.}.

The reason that both ideal hydrodynamics and parton transport 
could reproduce the same RHIC $v_2$ data is that
the earlier hydrodynamical calculations utilized a \emph{softer}
EOS ($\epsilon > 3p$) that incorporates the QGP phase transition,
which reduces elliptic flow.
Though dissipative effects for such realistic equations of state 
are not yet calculable,
there is no indication that they would be smaller than 
for the ideal gas studied here.

\begin{figure}[htpb]
\centerline{\epsfig{file=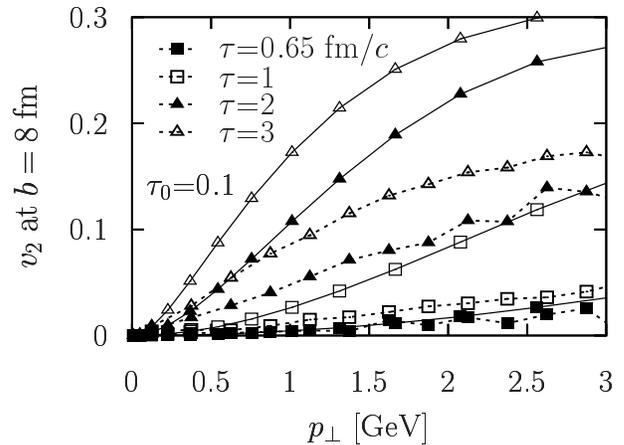,width=3.23in,height=2.38in,angle=0}}
\caption{Time evolution of $v_2(p_\perp)$ from ideal hydrodynamics (solid curves)
and transport theory for $\sigma_{gg}=47$ mb (dashed curves).}
\label{fig:2}
\end{figure}

The difference between transport and hydrodynamic elliptic flow is
established rather early during the evolution, as shown in Fig.~\ref{fig:2}
where we plot $v_2(p_\perp)$ computed over hypersurfaces of constant $\tau$.
Already by $\tau \approx 1$~fm/$c$,
the hydrodynamic anisotropy is a factor two or more above the transport,
especially at high $p_\perp$.
The transport $v_2$ builds up smoothly and at much the same
pace both at high and low $p_\perp$,
and by $\tau = 3$ fm$/c$ most ($\sim 80$\%) of 
the final anisotropy is there. This reinforces 
the short few-fm/$c$ timescales found for the $p_\perp$-integrated $v_2$ in 
\cite{Kolb:2000sd,ZPCv2}.
On the other hand the hydrodynamical development of anisotropy is more
$p_\perp$-dependent. Above $p_\perp = 1$ GeV most (85-90\%) of the anisotropy
is also developed at $\tau = 3$ fm/$c$, but at low $p_\perp$, e.g.\ at 
$p_\perp = 0.5$ GeV, anisotropy still increases by a factor of 1.5 before
freeze-out at $\tau=5.3$ fm/c.

A qualitative explanation for
the remarkable initial lag in the $v_2$ evolution for the transport
is that for an equilibrium initial condition
Eq.~(\ref{Boltzmann_eq}) in fact corresponds to {\em free streaming}
because the collision term vanishes exactly.
Momentum observables can start changing only after the system streams out of 
equilibrium.
In contrast, in hydrodynamics,
the initial pressure gradients induce changes in 
the flow pattern
and therefore immediately start to generate elliptic flow.

Because of scalings of the dynamical equations, the above results are
quite general.  For the transport, the solutions depend only on three
scales\cite{Molnar:v2}: $T_0$, a trivial scale that fixes the momentum
units; and two nontrivial scales $\sigma_{tot}dN/d\eta$ and
$R/\tau_0$.  For ideal hydrodynamics, one can similarly prove that
three scales apply: $T_0$ (momentum units), $dN/d\eta$ (linear scale
for particle number, e.g., the spectra), and $R/\tau_0$.  For example,
for twice larger $T_0$ and one-third the $dN/d\eta$, the transport
$v_2(p_\perp)$ solutions are the same provided $\sigma_{tot}$ is
increased three times and the $p_\perp$ axis is stretched by factor of
two, while the hydrodynamic $v_2(p_\perp)$ solutions can be obtained
via the stretching of $p_\perp$ axis alone 
if the freeze-out density is scaled in the same way as the initial density.

\begin{figure}[htpb]
\centerline{\epsfig{file=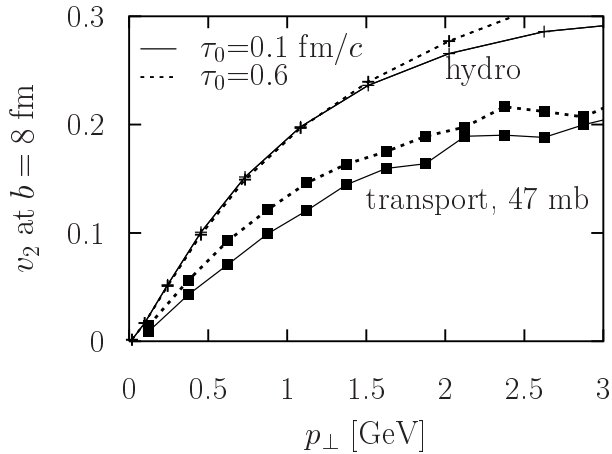,width=3.23in,height=2.38in,angle=0}}
\caption{Weak initialization time dependence of 
$v_2(p_\perp)$ from ideal hydrodynamics (pluses) and transport theory for
$\sigma_{gg}=47$ mb (boxes).}
\label{fig:3}
\end{figure}
Another important difference between the earlier
hydrodynamic and transport calculations 
was that the hydrodynamic evolution started at an assumed
\emph{thermalization} time $\tau_0 = 0.6$ fm/$c$, 
whereas the transport evolution started at a \emph{formation} 
time estimate of $\tau_0 = 0.1$ fm/$c$.
One might naively expect that the much smaller initial time for the transport
leads to more rapid departure from equilibrium and stronger
dissipative effects.
However, \emph{for both theories}
elliptic flow $v_2(p_\perp)$ is largely insensitive to the initialization time,
as shown in Fig.~\ref{fig:3}
where we compare the above mentioned results for $\tau_0 = 0.1$~fm$/c$ and
$T_0 = 700$ MeV with those for
$\tau_0 = 0.6$~fm$/c$ and $T_0 = 385$ MeV.
To approximately preserve typical particle momenta at freezeout,
we have rescaled the initial temperature according to a 1D
Bjorken expansion, under which $T\sim \tau^{-1/3}$.

The main reason for the insensitivity to initial time is
that the only relevant parameter
that differs between the two calculations, $\tau_0/R$,
is in both cases much smaller than one.
At early times when $(\tau-\tau_0) \ll R$,
the system essentially expands only longitudinally.
Therefore, between $\tau=0.1$ and 
$0.6$~fm$/c$ very little elliptic flow is generated, as can be seen in 
Fig.~\ref{fig:2}.
Our results reinforce similar findings for ideal hydrodynamics 
by Ref.~\cite{Ollitrault},
and also show that the elliptic flow pattern is largely insensitive to the 
initial time even in the presence of significant dissipation,
as long as $\tau_0 \ll R$.

{\em Conclusions.}  
In this paper we studied elliptic flow $v_2(p_\perp)$ using ideal
hydrodynamics and parton transport theory for an ideal gas of massless
partons.  From identical initial conditions and thermodynamic
properties, dissipation significantly reduced $v_2(p_\perp)$, even for
extreme $\sim 45$~mb elastic parton cross sections that are of the
order of hadronic cross sections. In addition, comparison to earlier
works shows that elliptic flow is more sensitive to dissipation than
the final transverse energy.

These results indicate that the large elliptic flow seen at RHIC can be 
interpreted in (at least) two ways:
either as i) a harder nuclear equation of state and strongly dissipative
evolution, or ii) a softer equation of state and
negligible dissipation which means that the system stays in
essentially perfect local kinetic equilibrium
via a mechanism whose microscopic origins we do not yet understand.

We emphasize that our results correspond to the simple ideal gas
equation of state. A detailed investigation of dissipation in nuclear
collisions will have to consider in the future a more realistic
nuclear equation of state that includes the hadronization phase
transition.

Finally, though our transport solutions do not reach the ideal
hydrodynamic limit, it would be very interesting to check how far they
are from the Navier-Stokes limit.

{\em Acknowledgments.}
Helpful discussions with M. Gyulassy and the hospitality of the Columbia 
University Nuclear Theory Group are gratefully acknowledged. 
P.H. acknowledges the hospitality of INT Seattle where part of this work was
done. This work was supported by DOE grants DE-FG02-01ER41190 and
DE-FG02-87ER40328.

\end{document}